\documentclass
[final,pre,amsfonts,amssymb,showkeys,preprint,noshowpacs,onecolumn,a4paper,groupedaddress,10pt,tightenlines]{revtex4}%
\usepackage{amsfonts}
\usepackage{amsmath}
\usepackage{amssymb}
\usepackage{graphicx}%
\providecommand{\U}[1]{\protect\rule{.1in}{.1in}}

\begin{document}
\title{The Human Group Optimizer (HGO): Mimicking the collective intelligence of
human groups as an optimization tool for combinatorial problems.}
\author{Ilario De Vincenzo$^{1}$, Ilaria Giannoccaro$^{1}$, Giuseppe Carbone$^{1,2,3}$}
\affiliation{$^{1}$Department of Mechanics, Mathematics and Management, Politecnico di
Bari, v.le Japigia 182, 70126 Bari - Italy}
\affiliation{$^{2}$Physics Department M. Merlin, CNR Institute for Photonics and
Nanotechnologies U.O.S. Bari via Amendola 173, 70126 Bari, Italy}
\affiliation{$^{3}$Department of Mechanical Engineering, Imperial College London, London,
South Kensington Campus, London SW7 2AZ, United Kingdom}
\begin{abstract}
A large number of optimization algorithms have been developed by researchers
to solve a variety of complex problems in operations management area. We
present a novel optimization algorithm belonging to the class of swarm
intelligence optimization methods. The algorithm mimics the decision making
process of human groups and exploits the dynamics of this process as an
optimization tool for combinatorial problems. In order to achieve this aim, a
continuous-time Markov process is proposed to describe the behavior of a
population of socially interacting agents, modelling how humans in a group
modify their opinions driven by self-interest and consensus seeking. As in the
case of a collection of spins, the dynamics of such a system is characterized
by a phase transition from low to high values of the overall consenus
(magnetization). We recognize this phase transition as being associated with
the emergence of a collective superior intelligence of the population. While
this state being active, a cooling schedule is applied to make agents closer
and closer to the optimal solution, while performing their random walk on the
fitness landscape. A comparison with simulated annealing as well as with a
multi-agent version of the simulated annealing is presented in terms of
efficacy in finding good solution on a NK - Kauffman landscape. In all cases
our method outperforms the others, particularly in presence of limited
knowledge of the agent.

\end{abstract}
\maketitle

\section{Introduction\label{introduction}}

Researchers have developed a large number of meta-heuristic algorithms
inspired by nature with the aim of solving combinatorial optimization
problems. A common way to classify them is to distinguish between trajectory
and population-based algorithms. Trajectory algorithms, such as Simulated
Annealing (SA) \cite{SA-Original} and Quantum Annealing \cite{QA, QA2},
describe a trajectory (usually a random walk) in the search space to reach the
solution. Population-based algorithms perform multiple search processes, each
of them carried out by a different agent. Population-based algorithms can be
further distinguished in two classes: i) Evolutionary algorithms and ii)
Swarm-based algorithms \cite{survey}. The evolutionary algorithms mimic the
processes of natural evolution, such as mutation, selection, and inheritance,
to identify the best solution. An example of such a class of algorithms is the
genetic algorithm \cite{evolutionary}. The swarm algorithms exploit the
collective intelligence of the social groups, such as flock of birds, ant
colonies, and schools of fish, in accomplishing different tasks. They include
the Ant Colony Optimization (ACO) \cite{DorigoOr0, DorigoOr1, DorigoOr2}, the
Particle Swarm Optimization \cite{PSO}, the Differential Evolution
\cite{Diff-Evo}, the Artificial Bee Colony \cite{ABC-0, ABC}, the Glowworm
Swarm Optimization \cite{GSO-1, GSO-2}, the Cuckoo Search Algorithm
\cite{CSA}, and very recently the Grey Wolf Optimizer \cite{GWO} and the Ant
Lion Optimizer \cite{ALO}.

All these algorithms have been successfully applied to solve production and
operation management problems. For example, Simulated Annealing has been
mainly employed to solve the traveling salesman problem \cite{golden},
scheduling problems \cite{ventura, melouk}, facility location and supply chain
design problems \cite{Arostegui, Jayaraman}. The Genetic Algorithms count a
larger number of applications compared to Simulated Annealing, even though the
wideness of the areas to which they have been applied is quite narrow
\cite{Chaudhry}. Aytug et al. \cite{Aytug} provide an interesting review of
the use of genetic algorithms for solving different types of operations
problems including production control, facility layout design, line balancing,
production planning, and supply chain management.

These last years have seen a huge growth of the applications of swarm-based
algorithms (in particular, ACO, bee colony, and swarm particle algorithms) in
operations management context \cite{REVIEW, REVIEW2, zhang, Hamta}. They share
remarkable features, such as decentralization, self-organization, autonomy,
flexibility, and robustness, which have been proven very useful to solve
complex operational tasks \cite{Ottino-Nature, Book-dorigo}. Applications of
ACO\ algorithm mainly concern the traveling salesman problem, scheduling,
vehicle routing, and sequential ordering \cite{dorigo1999}. More recently,
they have been also employed in supply chain contexts to solve
production-inventory problems \cite{Ferretti, Nia} and network design
\cite{Monkayo}.

In particular, these algorithms reproduce the collective decision making that
makes social groups superior in solving tasks compared to single individuals.
Agents (ants, bees, termites, fishes) make choices, pursuing their individual
goals (forage, survive, etc.) on the basis of their own knowledge and amount
of information (position, sight, etc.), and adapting their behavior to the
actions of the other agents. The group-living enables social interactions to
take place as a mechanism for knowledge and information sharing
\cite{KnowledgeTransf7, KnowledgeTransf1, KnowledgeTransf2, KnowledgeTransf3,
KnowledgeTransf4, KnowledgeTransf5, KnowledgeTransf6, KnowledgeTransf8, West,
knowF1, knowF2}. Even though the single agents may possess a limited
knowledge, and their actions are usually very simple, the collective behavior,
enabled by the social interactions, leads to the emergence of a superior
intelligence of the group.

In this paper we propose a novel swarm intelligence optimization algorithm to
solve complex combinatorial problems. The proposed algorithm is inspired by
the behavior of human groups and their ability to solve a very large variety
of complex problems, even when the individuals may be characterized by
cognitive limitations. Although it is widely recognized that human groups,
such as organizational teams, outperform single individuals in solving many
different tasks including new product development, R\&D activities, production
and marketing issues, literature is still lacking of optimization algorithms
inspired by the problem solving process of human groups. Similarly to other
social groups, human groups are collectively able, by exploiting the potential
of social interactions, to achieve much better performance than single
individuals can do. This specific ability of human groups has been defined as
group collective intelligence \cite{Pentland, Malone} that recently is
receiving a growing attention in the literature as to its antecedents and
proper measures \cite{Pentland, Malone}.

The proposed algorithm, hereafter referred to as Human Group Optimization
(HGO), is developed within the methodological framework recently proposed by
Carbone and Giannoccaro \cite{CG} to model the collective decision making of
human groups. This model captures the main drivers of the individual behavior
in groups, i.e., self-interest and consensus seeking, leading to the emergence
of collective intelligence. The group is conceived as a set of individuals
making choices based on rational calculation and self-interested motivations.
However, any decision made by the individual is also influenced by the social
relationships he/she has with the other group members. This social influence
pushes the individual to modify the choice he/she made, for the natural
tendency of humans to seek consensus and avoid conflict with people they
interact with \cite{consenso}. As a consequence, effective group decisions
spontaneously emerge as the result of the choices of multiple interacting individuals.

To test the ability of HGO algorithm, we compare its performance with those of
some benchmarks chosen among trajectory-based and population-based algorithms.
In particular, the HGO is compared with the Simulated Annealing (SA) and a
Multi Agent version of the Simulated Annealing (MASA). We also compare HGO
with a well-established swarm algorithm such as ACO in solving the traveling
salesman problem.

The paper is organized as follows. In Sec. \ref{the model} we briefly present
the decision making model of human groups, which the HGO algorithm relies on.
In Sec. \ref{swarm-intelligent} we discuss the conditions which lead to the
emergence of the group collective intelligence. Then, in Sec. \ref{HGO} we
present the Human Group Optimization (HGO) algorithm and its main features.
Sec. \ref{results} tests the HGO algorithm in solving $NP$ problems of
increasing complexity and compares it with the Simulated Annealing and a
Multi-Agent version of Simulated Annealing. In Sec. \ref{conclusion} we draw
the main conclusion and discuss future perspectives.

\section{The Decision Making Model of Human Groups\label{the model}}

Here we briefly summarize the decision making model presented in Ref.
\cite{CG}. We consider a human group made of $M$ socially interacting members,
which is assigned to accomplish a complex task. The task is modelled in terms
of $N$ binary decisions and the problem consists in solving a combinatorial
decision making problem by identifying the set of choices (configuration) with
the highest fitness, out of $2^{N}$ configurations.

As an example of application of the method, the fitness landscape, i.e., the
map of all configurations and associated fitness values, is generated
following the classical $NK$ procedure, where $N$ are the decisions and $K$
the interactions among them. Each decision $d_{i}$ of the vector $\mathbf{d}$
is a binary variable $d_{i}=\pm1$, $i=1,2,...,N$. Each vector $\mathbf{d}$ is
associated with a certain fitness value $V\left(  \mathbf{d}\right)  $
computed as the weighted sum of $N$ stochastic contributions $W_{j}\left(
d_{j},d_{1}^{j},d_{2}^{j},..,d_{K}^{j}\right)  $ that each decision leads to
the total fitness. The contributions $W_{j}\left(  d_{j},d_{1}^{j},d_{2}%
^{j},..,d_{K}^{j}\right)  $ depend on the value of the decision $d_{j}$ itself
and the values of other $K$ decisions $d_{i}^{j}$, $i=1,2,...,K$, and are
determined following the classical $NK$ procedure \cite{NK-model, NK-model1,
NK-model2}. The fitness function is then defined as
\begin{equation}
V\left(  \mathbf{d}\right)  =\frac{1}{N}\sum_{j=1}^{N}W_{j}\left(  d_{j}%
,d_{1}^{j},d_{2}^{j},..,d_{K}^{j}\right)  \label{fitness function}%
\end{equation}
The integer index $K=0,1,2,...,N-1$ corresponds to the number of interacting
decision variables, and tunes the complexity of the problem: increasing $K$
increases the complexity of the problem.

Individuals are characterized by cognitive limits, i.e. they posses a limited
knowledge. The level of knowledge of the $k$-th member of the group is
identified by the parameter $p\in\left[  0,1\right]  $, which is the
probability that each single member knows the contribution of the decision to
the total fitness.

Based on the level of knowledge, each member $k$ computes his/her own
perceived fitness (self-interest) as follows:%
\begin{equation}
V_{k}\left(  \mathbf{d}\right)  =\frac{\sum_{j=1}^{N}D_{kj}W_{j}\left(
d_{j},d_{1}^{j},d_{2}^{j},..,d_{K}^{j}\right)  }{\sum_{j=1}^{N}D_{kj}}.
\label{perceived fitness}%
\end{equation}
where $\mathbf{D}$ is the matrix whose elements $D_{kj}$ take the value $1$
with probability $p$ and $0$ probability $1-p$.

During the decision making process, each member of the group makes his/her
choices to improve the perceived fitness (self-interest) and to seek consensus
within the group. The dynamics is modelled by means of a continuos-time Markov
process where the state vector $\mathbf{s}$ of the system has $M\times N$
components $\mathbf{s=}\left(  s_{1},s_{2},...,s_{n}\right)  =\left(
\sigma_{1}^{1},\sigma_{1}^{2},...\sigma_{1}^{N},\sigma_{2}^{1},\sigma_{2}%
^{2},...\sigma_{2}^{N},...,\sigma_{M}^{1},\sigma_{M}^{2},...\sigma_{M}%
^{N}\right)  $. The variable $\sigma_{k}^{j}=\pm1$ is a binary variable
representing the opinion of the member $k$ on the decision $j$. The
probability $P\left(  \mathbf{s},t\right)  $ that at time $t$, the state
vector takes the value $\mathbf{s}$ out of $2^{N}$ possible states, satisfies
the master equation%
\begin{equation}
\frac{dP\left(  \mathbf{s},t\right)  }{dt}=-\sum_{l}w\left(  \mathbf{s}%
_{l}\rightarrow\mathbf{s}_{l}^{\prime}\right)  P\left(  \mathbf{s}%
_{l},t\right)  +\sum_{l}w\left(  \mathbf{s}_{l}^{\prime}\rightarrow
\mathbf{s}_{l}\right)  P\left(  \mathbf{s}_{l}^{\prime},t\right)
\label{Markov chain}%
\end{equation}
where $\mathbf{s}_{l}=\left(  s_{1},s_{2},.,s_{l}..,s_{n}\right)  $ and
$\mathbf{s}_{l}^{\prime}=\left(  s_{1},s_{2},.,-s_{l}..,s_{n}\right)  $. The
transition rate of the Markov chain (i.e. the probability per unit time that
the opinion $s_{l}$ flips to $-s_{l}$ while the others remain temporarily
fixed) is defined so as to be the product of the transition rate of the
Ising-Glauber dynamics \cite{Glauber}, which models the process of consensus
seeking to minimize the conflict level, and the Weidlich exponential rate
\cite{weidlich2, Sweitzer}, which models the self-interest behavior of the
agents:
\begin{equation}
w\left(  \mathbf{s}_{l}\rightarrow\mathbf{s}_{l}^{\prime}\right)  =\frac{1}%
{2}\left[  1-s_{l}\tanh\left(  \beta\frac{J}{\left\langle \kappa\right\rangle
}\sum_{h}A_{lh}s_{h}\right)  \right]  \exp\left\{  \beta^{\prime}\left[
\Delta V\left(  \mathbf{s}_{l}^{\prime},\mathbf{s}_{l}\right)  \right]
\right\}  \label{transition rates.}%
\end{equation}

In Eq. (\ref{transition rates.}) $A_{lh}$ are the elements of the adjacency
matrix, $J/\left\langle \kappa\right\rangle $ is the social interaction
strength and $\left\langle \kappa\right\rangle $ the mean degree of the
network of social interactions. The quantity $\beta$ is the inverse of the
social temperature that is a measure of the degree of confidence the members
have in the other judgement/opinion. Similarly, the quantity $\beta^{\prime}$
is related to the level of confidence the members have about their perceived
fitness (the higher $\beta^{\prime}$, the higher the confidence).

The pay-off function $\Delta V\left(  \mathbf{s}_{l}^{\prime},\mathbf{s}%
_{l}\right)  $ is simply the change of fitness perceived by the agent when its
opinion on the decision $j$ changes from $s_{l}$ to $-s_{l}$. The group
fitness value Eq. (\ref{fitness function}) is used as a measure of the
performance of the collective-decision making process. To calculate the group
fitness value, the vector $\mathbf{d=}\left(  d_{1},d_{2},...,d_{N}\right)  $
needs to be determined. To this end, consider the set of opinions $\left(
\sigma_{1}^{j},\sigma_{2}^{j},...,\sigma_{M}^{j}\right)  $ that the members of
the group have about the decision $j$, at time $t$. The decision $d_{j}$ is
obtained by employing the majority rule, i.e. we set:
\begin{equation}
d_{j}=\mathrm{sgn}\left(  M^{-1}\sum_{k}\sigma_{k}^{j}\right)  ,\qquad
j=1,2,...,N \label{majority rule}%
\end{equation}
If $M$ is even and in the case of a parity condition, $d_{j}$ is, instead,
uniformly chosen at random between the two possible values $\pm1$. The group
fitness is then calculated as $V\left[  \mathbf{d}\left(  t\right)  \right]  $
and the ensemble average $\left\langle V\left(  t\right)  \right\rangle $ is
then evaluated. The efficacy of the group in optimizing $\left\langle V\left(
t\right)  \right\rangle $ is then calculated as%
\begin{equation}
\eta\left(  t\right)  =\frac{\left\langle V\left(  t\right)  \right\rangle
-V_{\min}}{V_{\max}-V_{\min}} \label{efficacy}%
\end{equation}
where $V_{\max}$ and $V_{\min}$ are the maximal and minimal payoffs of the
fitness landscape. Note that $0\leq\eta\left(  t\right)  \leq1$.

The degree of consensus among the members is also computed. Following Ref.
\cite{CG} this is defined as:%
\begin{equation}
\chi\left(  t\right)  =\frac{1}{M^{2}N}\sum_{j=1}^{N}\sum_{kh=1}^{M}R_{hk}%
^{j}\left(  t\right)  \label{wholeconsensus}%
\end{equation}
where $R_{hk}^{j}\left(  t\right)  =\left\langle \sigma_{k}^{j}\left(
t\right)  \sigma_{h}^{j}\left(  t\right)  \right\rangle $. Observe that
$0\leq\chi\left(  t\right)  \leq1$.

\section{Criticality and swarm intelligence\label{swarm-intelligent}}

We refer to the case of a fully connected network of $M$ agents. We simulate
the Markov process by using the well-know stochastic simulation algorithm
proposed by Gillespie \cite{Gillespie1, Gillespie2}, see also Ref. \cite{CG}.

\begin{figure}[ptb]
\begin{center}
\includegraphics[
width=16.0cm
]{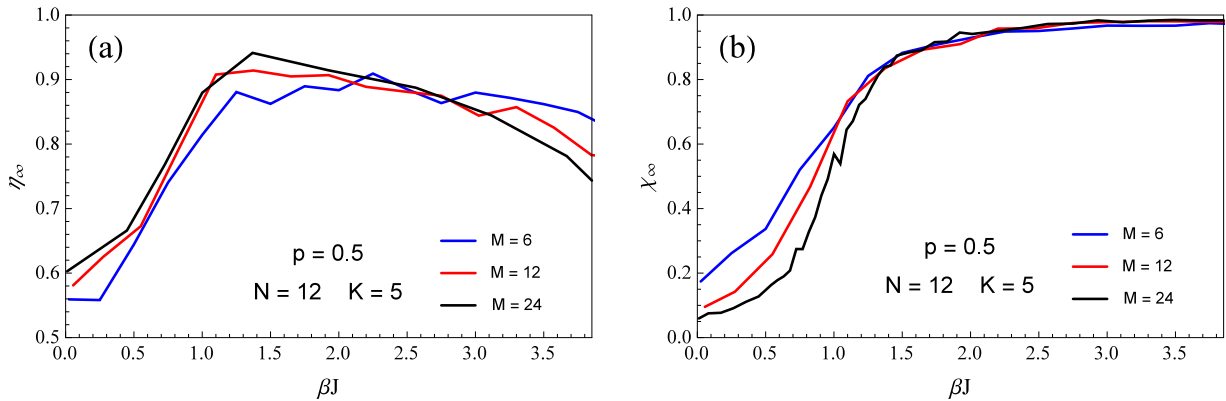}
\end{center}
\caption{The stationary values of the normalized averaged fitness
$\eta_{\infty}$ as a function of $\beta J$, (a); and of the statistically
averaged consensus $\chi_{\infty}$ as a function of $\beta J$, (b). Results
are presented for $p=0.5$, $K=5$ and for three different team sizes:
$M=6,12,24$.}%
\label{Criticality_and_Swarm_Intelligence}%
\end{figure}

Results are shown in Fig. \ref{Criticality_and_Swarm_Intelligence} where the
stationary values of efficacy $\eta_{\infty}=\eta\left(  t\rightarrow
\infty\right)  $ and the degree of consensus $\chi_{\infty}=\chi\left(
t\rightarrow\infty\right)  $ are reported as a function of the quantity $\beta
J$ for different group sizes $M=6,12,24$, $N=12$, $K=5$, $\beta^{\prime}=10$
and for an average level of knowledge $p=0.5$.\ Results clearly show that a critical threshold value of $\beta
J$ exists at which both consensus and payoff have a sharp and concurrent
increase. Notably, the transition from low to high payoff, accompanied by an
analogous transition from low to high consensus, becomes sharper as the group
size $M$ is increased. However, in all cases, given $\beta^{\prime}=10$, the
transition occurs for $\left(  \beta J\right)  _{C}\approx1$. Interestingly
this threshold value actually corresponds to the critical ordering transition
of the Ising model on a complete graph, in the thermodynamic limit of large
$M$. This result can be obtained by using the findings by Vespignani and
Mendes \cite{vespignani, Mendes}, who independently demonstrated that for
general graphs the critical transition of the Ising model occurs at%
\begin{equation}
\left(  \beta\frac{J}{\left\langle \kappa\right\rangle }\right)  _{C}%
=-\frac{1}{2}\log\left(  1-2\frac{\left\langle \kappa\right\rangle
}{\left\langle \kappa^{2}\right\rangle }\right)  \label{criticalthreshold}%
\end{equation}
Thus, considering that for complete graph $\left\langle \kappa\right\rangle
=M-1$, $\left\langle \kappa^{2}\right\rangle =\left(  M-1\right)  ^{2}$ and
that $M$ is large, expanding Eq. (\ref{criticalthreshold}) at first order in
$\left\langle \kappa\right\rangle /\left\langle \kappa^{2}\right\rangle $
gives $\left(  \beta J\right)  _{C}=1$. However, calculations shows that
increasing $\beta^{\prime}$ above $10$ makes the transition occur at values of
$\beta J$ smaller than one.

Based on these outcomes, the condition that leads to the emergence of the
collective intelligence (i.e. high value of efficacy) is simply identified by
the critical transition point at which consensus sets in. At this value of
consensus, a fully exploitation of the potential of social interactions is
obtained. In fact, as soon as the critical threshold value of $\left(  \beta
J\right)  _{C}$ is reached, the agent with limited knowledge, driven by the
social interactions, will make good choices following those group members, who
have higher knowledge about the problem.\begin{figure}[ptb]
\begin{center}
\includegraphics[
width=16.0cm
]{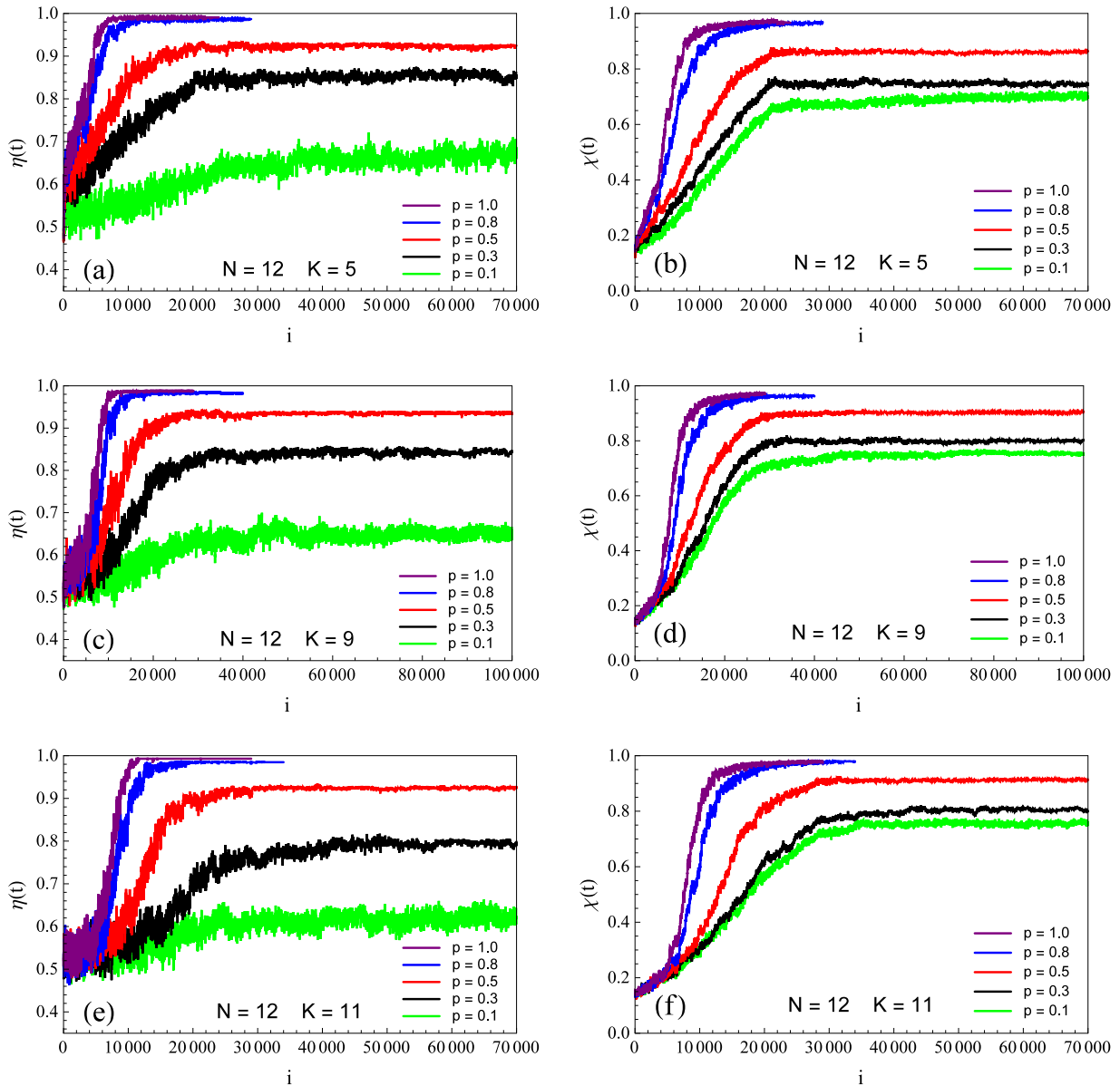}
\end{center}
\caption{The time-evolution of the efficacy $\eta\left(  t\right)  $ and
degree of consensus $\chi\left(  t\right)  $ for $p=0.1,0.3,0.5,0.8,1.0$, and
$K=5,9,11$.}%
\label{Figure2}%
\end{figure}\begin{figure}[ptb]
\begin{center}
\includegraphics[
width=16.0cm
]{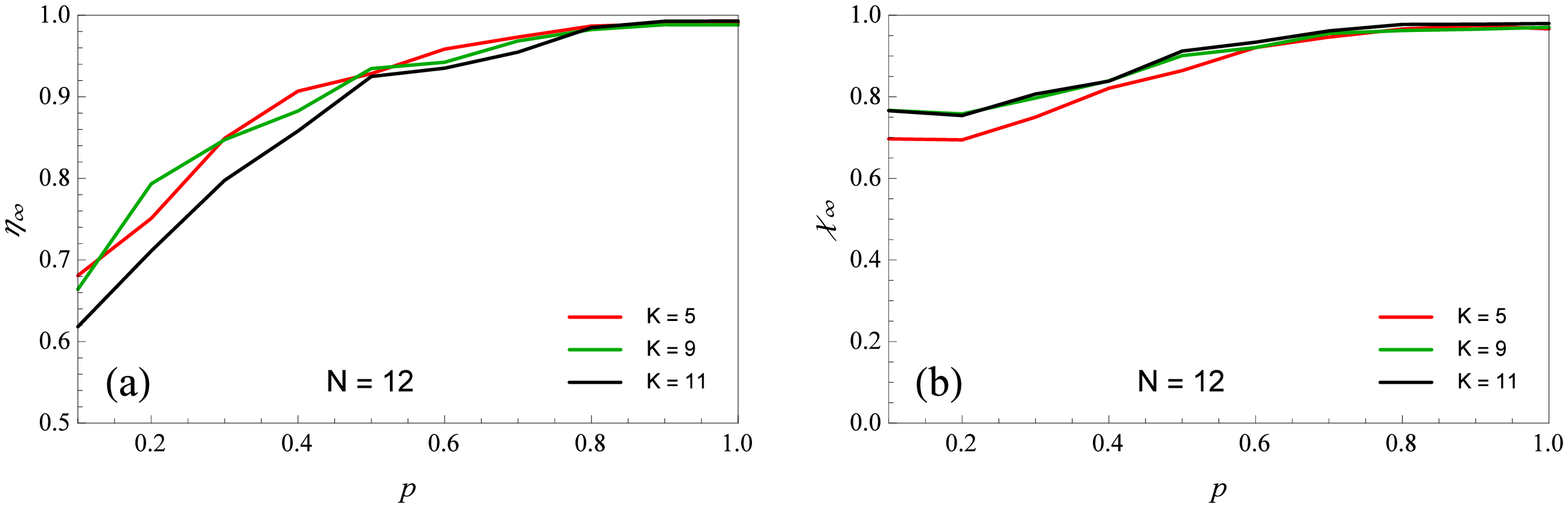}
\end{center}
\caption{The steady-state efficacy $\eta_{\infty}$ (a), and degree of
consensus $\chi_{\infty}$ (b), as a function of the knowledge level $p$, for
$N=12$, $M=7$, and $K=5,9,11$.}%
\label{Figure3}%
\end{figure}\begin{figure}[ptb]
\begin{center}
\includegraphics[
width=16.0cm
]{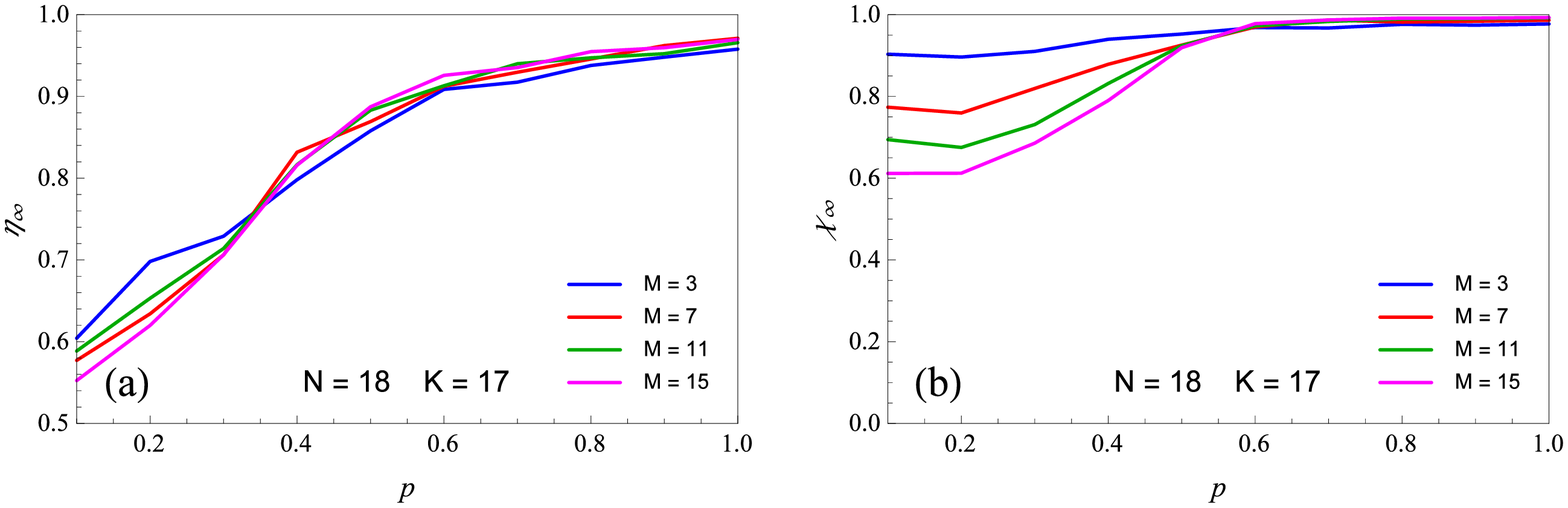}
\end{center}
\caption{The steady-state efficacy $\eta_{\infty}$ (a), and degree of
consensus $\chi_{\infty}$ (b), as a function of the knowledge level $p$, for
$N=18$, $K=17$, and $M=3,7,11,15$.}%
\label{Figure4}%
\end{figure}\begin{figure}[ptb]
\begin{center}
\includegraphics[
width=16.0cm
]{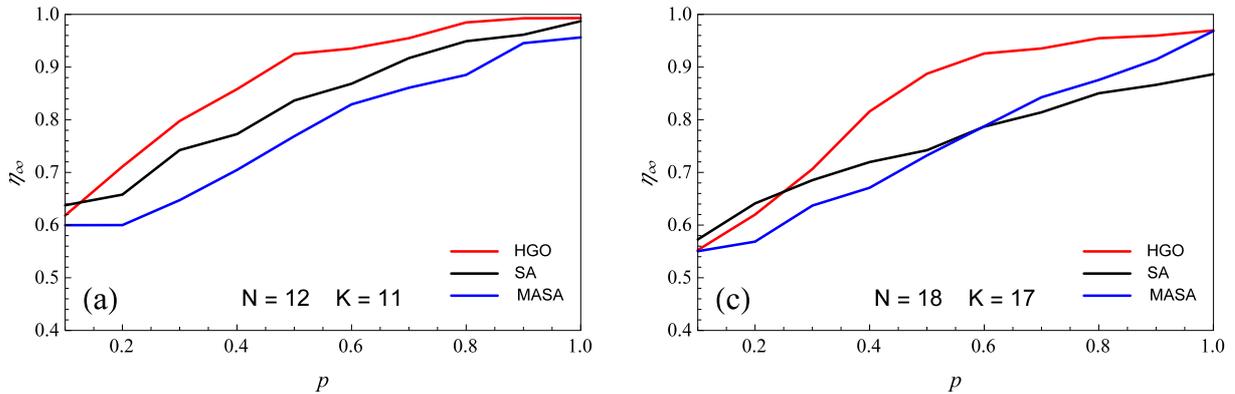}
\end{center}
\caption{A comparison between the proposed HGO, SA, and MASA, in terms of
steady-state efficacy $\eta_{\infty}$ as a function of the knowledge level
$p$, for $N=12$, $K=11$, (a) and $N=18$, $K=17$, (b).}%
\label{Figure5}%
\end{figure}

\section{The Human Group Optimization algorithm \label{HGO}}

In this section we design the HGO algorithm exploiting the collective
intelligence property of the decision making process to solve combinatorial
problems. To this aim, we emulate the process followed to design the Simulated
Annealing algorithm \cite{SA-Original}. We first observe that the Markov
process defined in Eq. (\ref{Markov chain}) with transitions rates
Eq.(\ref{transition rates.}) converges to the stationary probability
distribution \cite{CG}%

\begin{equation}
P_{0}\left(  \mathbf{s}_{l}\right)  =\frac{\exp\left[  -\beta E\left(
\mathbf{s}_{l}\right)  +2\beta^{\prime}\bar{V}\left(  \mathbf{s}_{l}\right)
\right]  }{\sum_{k}\exp\left[  -\beta E\left(  \mathbf{s}_{k}\right)
+2\beta^{\prime}\bar{V}\left(  \mathbf{s}_{k}\right)  \right]  }
\label{Boltzmann}%
\end{equation}
where the total level of conflict is $E\left(  \mathbf{s}\right)
=-0.5\left\langle \kappa\right\rangle ^{-1}J\sum_{ij}A_{ij}s_{i}s_{j}$. Eq.
(\ref{Boltzmann}) is a Boltzmann distribution with effective energy
\begin{equation}
E_{\mathrm{eff}}\left(  \mathbf{s}_{l}\right)  =-\bar{V}\left(  \mathbf{s}%
_{l}\right)  +\alpha E\left(  \mathbf{s}_{l}\right)
\end{equation}
where $\alpha=\beta/\left(  2\beta^{\prime}\right)  $. We then make the
parameters $\beta J$ and $\beta^{\prime}$ change during the process as
follows:%
\begin{align}
\beta^{\prime}  &  =\beta_{0}^{\prime}\log\left(  i+1\right) \\
\beta J  &  =\min\left\{  \mu(i-1),\left(  \beta J\right)  _{C}\right\}
\nonumber
\end{align}
where $i\ $is the time iterator, $\mu$ is chosen by the user, and $\beta
_{0}^{^{\prime}}$ is set according to Ref. \cite{initial temperature}. These
requirements assure that the critical transition to the collective
intelligence state is completed during the process, and that $\alpha$ vanishes
in the long term limit so as to allow $E_{\mathrm{eff}}\left(  \mathbf{s}%
_{l}\right)  \rightarrow-\bar{V}\left(  \mathbf{s}_{l}\right)  $. Note that,
when individuals possess complete knowledge ($p=1$), the latter condition,
akin the Simulated Annealing, makes the proposed algorithm converge in
probability to the optimum of $V\left(  \mathbf{d}\right)  $
\cite{Stuart-SA-Convergence, Hajeck}.

Also observe that the choice $\beta J=0$ identifies an optimization algorithm
very closely related to the Simulated Annealing, except that the fitness
landscape is explored by $M$ non-interacting agents. Hereafter, this algorithm
will be referred to as Multi Agent Simulated Annealing (MASA). Observe that
MASA is characterized by the absence of social interactions among the agents,
and, as such, it is unable to exploit the swarm intelligence of the group.

\section{Simulation and results \label{results}}

In this section we first analyze the performance of the HGO algorithm for the
case of a $NK$ landscape with $N=12$ and $K$ ranging from $5\ $to $11$. A much
more complex case is also analyzed with $N=18$ and $K=17$. We also investigate
the effect of the size of the group on the performance of the HGO, by making
$M$ range from $3$ to $15$. We assume that the network of social interaction
among the $M$ agents is described by a complete graph. Then, we compare HGO
with Simulated Annealing (SA) and Multi Agent Simulated Annealing (MASA) for
the case of $N=12$ and $K=11$, $N=18$ and $K=17$. Comparison is carried out at
increasing levels of knowledge $p$.

In all simulations each stochastic process is simulated by generating $50$
different realizations and the ensemble average of the results is then
calculated. The simulation is stopped at steady-state, i.e., when changes in
the time-averages of consensus and pay-off over consecutive time intervals of
a given length is sufficiently small.

HGO performance in solving complex problems

In Fig. \ref{Figure2} the time-evolution ($i$ is the time iterator) of the HGO
performance are reported for $N=12$, $K=5,9,11$, and different levels of
knowledge $p$ ranging from $0.1$ to $1$. We observe that independently of the
complexity level $K\ $and level of knowledge $p$, the increase of $\eta\left(
t\right)  $ is always accompanied by simultaneously increase of $\chi\left(
t\right)  $. This confirms that the transition to swarm intelligence always
occurs, and that this condition is necessary to guarantee high performance of
the HGO algorithm. The complexity parameter $K$ only marginally affects the
performance of the method. The level of knowledge $p$ of the agents, instead,
strongly affects the performance of the optimization algorithm. However, high
efficacy can be achieved already at moderate levels of knowledge: $p=0.5$
determines a final efficacy close to $0.9$. This result is also shown in Fig.
\ref{Figure3}, where the steady-state values of the efficacy $\eta_{\infty}$
[Fig. \ref{Figure3}(a)], and consensus $\chi_{\infty}$ [Fig. \ref{Figure3}(b)]
are plotted as a function of the level of knowledge $p$, for $K=5,9,11$.

Figure \ref{Figure4} shows $\eta_{\infty}\ $and $\chi_{\infty}$ as a function
of $p$ for different group sizes $M=3,7,11,15$, $N=18$ and $K=17$. We note
that, for $p>0.3$, increasing $M$ slightly improves the outcome of the process
i.e. the efficacy of the optimization method [Fig. \ref{Figure4}(a)]. In all
cases we still notice that the best results are obtained for $p>0.5$ which
seems to be a threshold value that must be exceeded to guarantee a high degree
of consensus $\chi_{\infty}$ among the agents [Fig. \ref{Figure4}(b)], and, in
turn, high fitness values [Fig. \ref{Figure4}(a)].

Fig. \ref{Figure5} compares the HGO with SA and MASA. Results are shown for
$N=12$, $K=11$, [Fig. \ref{Figure5}(a)] and $N=18$, $K=17$ [Fig.\ref{Figure5}%
(b)], with $p$ ranging from 0 to 1. In the case of HGO and MASA, we use $M=7$.
In all cases the HGO algorithm outperforms the other methods. However, the
most significant differences are observed in the case of limited knowledge of
the agents. In these situations HGO strongly outperforms the Simulated
Annealing and the Multi-agent Simulated Annealing. In this case, the social
interaction among the agents pushes individuals, who do not have knowledge
about a certain decision, to make good choice following the decisions of the
agents who instead know the influence of the decision on the fitness values,
thus making the entire group perform much better compared to the case of non
socially interacting members.

\section{Conclusions \label{conclusion}}

In this paper we proposed a novel swarm-based optimization algorithm mimicking
the collective decision-making behavior of human groups. This algorithm, which
we termed Human Group Optimization (HGO), describes the decision process of
the agents in terms of a time-continuous Markov chain, where the transition
rates are defined so as to capture the effect of the self-interest, which
pushes each single agent to increase the perceived fitness, and of social
interactions, which stimulate member to seek consensus with the other members
of the group. The Markov chain is, then, characterized by a couple of
parameters that, likewise the Simulated Annealing, are subjected to a specific
cooling schedule that in the long-time limit makes the system converge in
probability to the optimal value. The choice of the parameters is made in
order to guarantee the transition to a consensus state at which the group of
agents shows a very high degree of collective intelligence. While being in
this state, the agents explore the landscape by sharing information and
knowledge through social interactions, so as to achieve very good solutions
even in the case of a limited knowledge.

To test the proposed HGO algorithm, we considered the hard-$NP$ problem of
finding the optimum on $NK$ fitness landscape and compared the methodology
with other well established algorithms as the Simulated Annealing and a
multi-agent version of it. In all cases the HGO has been shown to
significantly outperform the other two algorithms, especially under limited
knowledge conditions. Summarizing, our algorithm presents several advantages
that make it very suitable to solve complex operation management problems. It
is flexible because it can be applied to almost any combinatorial problem by
identifying the number of decisions the agents should make. However, its most
attractive feature relies in its ability to identify very good solutions, even
in presence of partial knowledge of the agents. For this reason it appears
very promising for applications in distributed decision making contexts such
as supply chains. Furthermore, while the vast majority of swarm intelligent
algorithms, mimicking the behavior of social groups like insects and animals,
are based on the mechanism of the stigmergy, our algorithm introduces a
mechanism based on the direct communication among individuals, which is a more
powerful and effective way to achieve coordination. Under this perspective,
the proposed code is novel and unique within the class of swarm intelligent
optimization codes.

We recognize that this first version of the algorithm could be further
improved in future research by identifying better cooling schedules. The
algorithm could be also fine-tuned to solve specific operations management
problems characterized by distributed decision making and information
asymmetry, such as multi-stage production scheduling, location routing
problem, supply chain inventory problem, just to name a few. Additional
numerical tests and theoretical investigation, not in the scope of present
study, are however needed to quantify pros and cons.

\end{document}